\documentclass[11pt]{article}

\usepackage{fullpage}
\usepackage{libertine}

\usepackage{color}

\usepackage{graphicx}
\usepackage{subcaption}
\usepackage{mathtools}
\usepackage{amssymb,amsmath,amsfonts,amstext,amsthm}

\newtheorem{thm}{Theorem}
\newtheorem{prop}{Proposition}
\newtheorem{cor}{Corollary}

\theoremstyle{definition}

\newcommand{\R}{\mathbb{R}}
\newcommand{\N}{\mathbb{N}}

\newcommand{\calA}{\mathcal{A}}

\newcommand{\calR}{\mathcal{R}}
\newcommand{\calC}{\mathcal{C}}

\newcommand{\charger}{\text{charger}}

\newcommand{\inside}{\text{in}}

\newcommand{\norm}[1]{\left\lVert #1 \right\rVert}

\newcommand{\remove}[1]{}

\usepackage{authblk}
\usepackage{natbib}

\begin{document}
\title{\bf Adaptive Wireless Power Transfer in \\ Mobile Ad Hoc Networks\thanks{A preliminary version of this paper entitled ``Mobility-Aware, Adaptive Algorithms for Wireless Power Transfer in Ad Hoc Networks'' appears in {\em Proceedings of the 14th International Symposium on Algorithms and Experiments for Wireless Sensor Networks (ALGOSENSORS), 2018}. In this full version, we demonstrate the need for adaptiveness not only theoretically but also experimentally. Further, we have also considered the fundamental special case of the MNL optimization problem with a single agent and on-off charger (1-MNLb). Finally, we have also implemented further simulations to showcase the scalability of our proposed algorithms in regards to some of the setting parameters. This work has been partially supported by the Greek State Scholarships Foundation (IKY), and by a PhD scholarship from the Onassis Foundation. The third author would like to thank Ioannis Caragiannis for fruitful discussions at early stages of this work.}}

\author{Adelina Madhja \quad Sotiris Nikoletseas}
\affil{\em Department of Computer Engineering and Informatics, University of Patras, Greece \\
\em Computer Technology Institute and Press ``Diophantus" (CTI), Greece}

\author{Alexandros A. Voudouris}
\affil{\em Department of Computer Science, University of Oxford, UK}

\date{}

\maketitle   

\begin{abstract}
We investigate the interesting impact of mobility on the problem of efficient wireless power transfer in ad hoc networks.  We consider a set of mobile agents (consuming energy to perform certain sensing and communication tasks), and a single static charger (with finite energy) which can recharge the agents when they get in its range. In particular, we focus on the problem of efficiently computing the appropriate range of the charger with the goal of prolonging the network lifetime. We first demonstrate (under the realistic assumption of fixed energy supplies) the limitations of any fixed charging range and, therefore, the need for (and power of) a dynamic selection of the charging range, by adapting to the behavior of the mobile agents which is revealed in an online manner. We investigate the complexity of optimizing the selection of such an adaptive charging range, by showing that two simplified offline optimization problems (closely related to the online one) are NP-hard. To effectively address the involved performance trade-offs, we finally present a variety of adaptive heuristics, assuming different levels of agent information regarding their mobility and energy.
\end{abstract}

\section{Introduction}\label{sec:intro}

Over the last decade, the continuously increasing development and excessive use of energy-hungry mobile devices (like smartphones, tablets, or even electric vehicles; see \citep{BKMZZK16,LM15}) in ad hoc networks, has given rise to the problem of efficient power management under various objectives. A viable solution to this critical problem, that has been extensively studied in the recent related literature due to its efficiency and wide applicability, is the Wireless Power Transfer (WPT) technology using magnetic resonant coupling~\citep{K+83} combined together with ultra-fast rechargeable batteries \citep{battery}. By exploiting such a technology, it is possible to recharge the network devices as required and prolong their lifetime (for instance, see the papers \citep{cannon09, IH11, WCS04} for some more recent advances on WPT using magnetic resonant coupling).

In rechargeable ad hoc networks, there are two main types of entities that are distributed in the network area, called {\em chargers} and {\em agents}, respectively. Usually, a charger is a special device that has high energy supplies and acts as a transmitter, while an agent has significantly lower battery capacity and acts as a receiver. The charger is responsible for the energy management in the network, by effectively transferring parts of its energy to the agents. In contrast, the agents are the actual network devices which consume energy while performing  communication and sensing tasks (like collecting and routing data) and are, therefore, in need of energy replenishment to sustain their normal operation.  

There are generally many different assumptions regarding the charging process, whether there is a single or multiple chargers that are mobile or not, as well as the information that is available about the energy levels and the locations of the (possibly mobile) agents. As the survey of all these different settings are not the main focus of this paper, we refer the interested reader to the book~\citep{nikole}. 

\subsection{Our contribution}
We consider ad hoc networks that consist of mobile agents and a single static charger. The agents move around in the network area and consume energy for communication purposes. 
The charger is assumed to have initial finite energy that can be used to replenish the battery of the agents that get in its charging range. The finite energy assumption here is well motivated in scenarios where we would like to cover isolated areas (for instance, mountains where people go hiking) and there are simply no wired sources capable to provide unlimited energy to the charger. See Section~\ref{sec:model} for a description of our model. 

As the mobility and energy consumption characteristics of the agents become available online, the charger must respond to the behavior of the agents by dynamically changing its transmission power which, in turn, defines the charging range. The main objective of this adaptive selection of the charging range is to extend the network lifetime, which can be defined as the time period during which there is at least one agent with non-zero energy or the time period during which a percentage of agents have non-zero energy; of course, this is not the only objective that one may be interested in. To the best of our knowledge, this is the first paper that systematically studies the setting where the charging range is dynamically selected adaptively to the agents status.

We theoretically and experimentally showcase the need for adaptiveness. In particular, for every possible fixed range that the charger may have, we identify worst-case scenarios where there is always an adaptive solution that performs better (see Section~\ref{sec:need-for-adaptiveness}).
In addition, we define two simplified offline optimization problems that are closely related to the online multi-objective one, and prove their computational intractability (see Section~\ref{sec:hardness}). Furthermore, we design three adaptive algorithms and compare them to each other in terms of various metrics using a non-trivial simulation setup, where we consider probability distributions over randomized mobility and energy consumption scenarios that are designed to test our methods in highly heterogeneous instances (see Section~\ref{sec:algorithms}).

\subsection{Related work}
Mobility in ad hoc networks has been thoroughly studied and many models have been proposed over the years. Generally, such mobility models assume that the agents perform different kinds of random walks that may depend on many different parameters~\citep{BRS03,CBD02}, and even be influenced by social network attributes that attempt to capture human behavior~\citep{HMDP17,MHM04,VY14}). In this work, we slightly deviate from previous work and adopt a mobility model that allows us to construct many interesting mobility patterns for the agents, that can also simulate cases where human mobility may be arbitrary, greedy or even irrational.

Recharging in mobile ad hoc networks has been the focus of many research papers. Indicatively, \citet{NRST17} considered mobile ad hoc networks with multiple static chargers of finite energy supplies, and evaluated (using real devices) two algorithms that decide which chargers must be active during each round, in order to maximize charging efficiency and achieve energy balance, respectively. \citet{ABER15} also considered mobile ad hoc networks, with the difference that there exists a single mobile charger that has infinite energy and traverses the network in order to recharge the agents as needed. They focused on designing optimal traversal strategies for the mobile charger with the goal of prolonging the network lifetime. 

\citet{HCJYXS13} studied the energy provisioning problem to minimize the number of chargers and compute where they should be located in the network area so that all (possibly) mobile agents are always active (have or get enough energy to complete their tasks). By taking into account an agent's velocity and battery capacity, \citet{DC4W15} showed that the agent's continuous operation cannot be guaranteed, and introduced the Quality of Energy Provisioning (QoEP) metric to characterize the expected time that the agent is actually active.

\citet{DLCWHLM17} considered static networks and studied the safe charging problem to maximize charging utility, while simultaneously ensuring that the electromagnetic radiation (EMR) does not exceed a threshold value at any point of the network. In~\citep{DLCWH14}, the authors studied a variation, where the power of each charger can be adjusted once at the beginning of time. \citet{radiation} studied the low radiation efficient wireless charging problem as well, but they defined a different charging model that takes into account hardware constraints for the chargers and the agents. The last two papers are the most related ones to ours, in the sense that the power of each charger is adjustable. However, observe that since the agents are static in both models considered in \citep{DLCWH14,radiation}, each charger adjusts its power only once, at the beginning of the time horizon. In contrast, the power of the charger in our setting constantly changes over time, adaptively to the behavior of the mobile agents which is revealed in an online manner. Therefore, even though our setting and that of \citep{DLCWH14,radiation} are seemingly similar, they are fundamentally different and uncomparable to each other. 

On a more practical level, the idea of adapting the wireless power of the charger has been studied before in terms of carefully adjusting either its frequency or its circuit components. For instance, \citet{SHMB08} presented a method to regulate the power that is transferred over a wireless link, by adjusting the resonant operating frequency of the charger. More recently, \citet{CKH17} analyzed how the power transfer performance is impacted by the load, distance, and coil alignment of the network devices, and introduced a cognitive wireless charger, which adaptively controls the operating frequency in real-time using implicit feedback from sensing for optimal operations. \citet{WS12} considered adaptive impedance matching network topologies, which can automatically change the input and output impedances in order to maintain maximum wireless power transfer efficiency at a single frequency, and presented algorithms to efficiently determine the component values. Our work differs significantly from these studies as we focus more on the conceptual and algorithmic level of adaptive charging power in order to showcase its merits in improving the network lifetime, and study the hardness of coming up with efficient adaptive solutions.

There are several studies that deviate from the above modeling assumptions. In particular, \citet{ZWL15} introduced the notion of collaborative charging, where the chargers are able to transfer energy to each other as well. This feature was extended by \citet{MNR16} in a hierarchical structure. Furthermore, recent studies do not even use chargers, but they assume that the agents themselves are able to both receive and send power wirelessly~\citep{MNRT16,MNTV18,NRR17}. Another research direction deals with the simultaneous energy transfer and data collection by the charger (e.g. \citep{ZLY16}). In this setting, practically, the charger acts as an energy transmitter as well as a sink.

\section{Model definition}\label{sec:model}
There are $n$ agents that move around in a bounded network area $\calA$, and a single static charger that is positioned at the center of $\calA$.\footnote{Actually, as we will see later in much more detail, the position of the charger does not essentially affect the selection of the charging range. Since we assume the existence of a single charger, the distance of the agents from the charger is the quantity that plays the most crucial role in the determination of the charging range. However, for settings with more than two chargers, the positions of the chargers can affect the charging performance, depending on the underlying charging model.} For simplicity, we assume that $\calA$ is represented by a rectangle defined by the points $(0,0)$ and $(x_{\max},y_{\max})$ on the Euclidean space. Hence, the position $p_\charger$ of the charger is given by the coordinates $(\frac{1}{2}x_{\max},\frac{1}{2}y_{\max})$. Further, we assume that there is a discrete time horizon $T \in \N_{\geq 0}$ consisting of a number of distinct rounds each of which runs for a constant period of time $\tau$. For every agent $i$, we denote by $p_i(t) = (x_i(t),y_i(t)) \in \calA$ its position at the beginning of round $t$. The positions of the agents are updated as they move around in $\calA$. For the charger, we denote by $R(t) \in [R_{\min},R_{\max}]$ its range during round $t$. $R(t)$ is decided by the transmission power of the charger and defines a circle of radius $R(t)$ around $p_\charger$; let $\calC_{R(t)} \subseteq \calA$ denote this circle on the plane. All agents that pass through $\calC_{R(t)}$ during round $t$ can get recharged (if they need to).

\subsection{Energy model}
Let $E_i(t)$ be the energy of agent $i$ at the beginning of round $t$. All agents have the same battery characteristics in the sense that they have the same battery capacity $B$. We assume that initially all agents are fully charged, i.e., $E_i(1) = B$ for every agent $i$.
During round $t$, each agent $i$ consumes an amount of energy $E_i^c(t)$ for communication purposes which depends on random sensing and routing events. Since the thorough study of such events are out of the scope of this paper, following previous work (e.g., see \citep{ABER15}), we simply assume that $E_i^c(t)$ follows a poisson probability distribution with expected value $\gamma_i \in [\gamma_{\min}^i, \gamma_{\max}^i]$. The energy of agent $i$ at the beginning of the next round $t+1$ (assuming no recharging takes place), is equal to 
$$E_i(t+1) = \max\left\{ 0, E_i(t) - E_i^c(t) \right \}.$$
We remark that the agents are assumed to {\em not} consume any energy due to movement as the necessary energy can be supplied by different sources. For example, in any crowdsensing scenario it is supplied by the humans that carry around their smart devices.  

\subsection{Charging model}
\begin{figure}[t]
\centering
\includegraphics[scale=0.3]{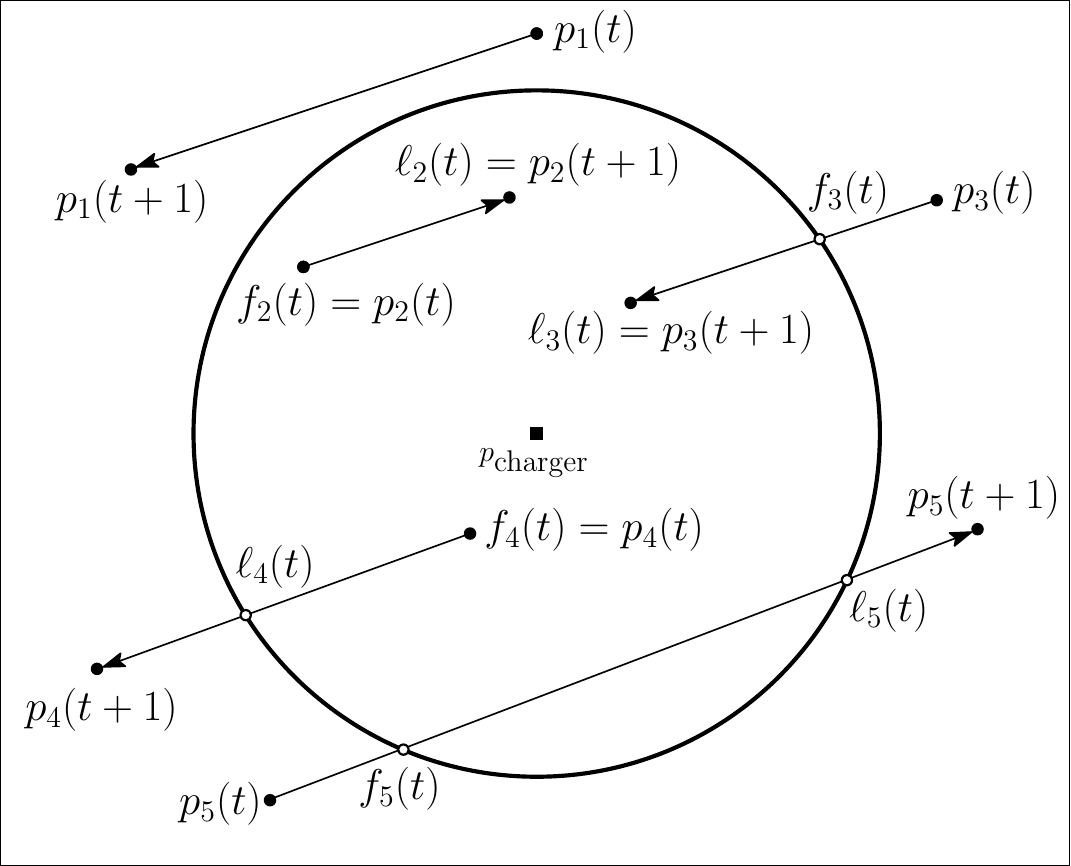}
\caption{An example of all possible cases regarding the relation between the line along which an agent may travel and $\calC_{R(t)}$. Here, agent $1$ does not get in range and, hence, $f_1(t)$ and $\ell_1(t)$ are undefined. Agent $2$ starts and ends in range, agent $3$ starts out of range but ends up inside, agent $4$ starts inside but ends up out of range, and finally agent $5$ travels through $\calC_{R(t)}$.}
\label{fig:in-range-cases}
\end{figure}

Let $E_\charger(t)$ denote the energy that the charger has at the beginning of round $t$. We assume that the charger initially has some {\em finite} amount of energy $E_\charger(1) = C$ that can be used to replenish the energy that the agents consume. 
In particular, if the charger has the appropriate amount of energy, then all agents that get in its range receive a positive amount of energy. Let $f_i(t)$ and $\ell_i(t)$ be the first and last position of agent $i$ that are in range. These may or may not be defined depending on whether the agent travels through $\calC_{R(t)}$ or not; Figure~\ref{fig:in-range-cases} depicts an example of all possible cases about the relations between $p_i(t)$, $p_i(t+1)$, $f_i(t)$ and $\ell_i(t)$. The time that agent $i$ spends in the charger's range is then equal to
\begin{align*}
T_i^\inside(t) = 
\begin{cases}
\frac{\norm{f_i(t)-\ell_i(t)}}{v_i(t)}, & \text{if } f_i(t) \neq \ell_i(t), v_i(t) \neq 0\\
\tau, & \text{if } f_i(t)=\ell_i(t), v_i(t)=0 \\
0, & \text{otherwise},
\end{cases}
\end{align*}
where $\norm{f_i(t)-\ell_i(t)}$ denotes the Euclidean distance between points $f_i(t)$ and $\ell_i(t)$. In order to keep our setting and discussion simple (as well as focus on the core of our problem), we assume that agent $i$ receives energy according to a simplified version of the well-known Friis transmission equation; see Section~\ref{sec:conclusion} for a more extensive discussion on charging models. In particular, 
\begin{align}\label{eq:received-energy}
E_i^r(t) = \frac{\alpha \cdot R(t)^2 \cdot T_i^\inside(t)}{(\norm{p_{\charger}-f_i(t)} + \beta)^2},
\end{align}
where $\alpha$ and $\beta$ are environmental and technological constants.  The energy of agent $i$ at the beginning of round $t+1$ (accounting for both energy consumption and recharging), is equal to 
$$E_i(t+1) = \min\left\{B,  \max\{0, E_i(t) - E_i^c(t) + E_i^r(t) \right\}\}.$$
Observe that the amount of energy that the agent receives must respect its battery limit. Of course, the energy of the charger is also decreased accordingly.

\section{The need for adaptiveness}\label{sec:need-for-adaptiveness}
Here, we aim to justify the need for algorithms that dynamically change the charging  range over time by adapting to the behavior of the agents. The simplest algorithm that one could  come up with, is to have the range {\em fixed} during the whole period of time; this is the typical algorithm that has been used in most of the related literature so far. However, observe that there are essentially infinitely many different fixed values. Therefore, finding the one that works efficiently (with respect to the various objectives that we could be interested in) for {\em every} possible instance is improbable. In fact, in the following we will prove that this is actually impossible.

\subsection{Theoretical justification}\label{sec:need-for-adaptiveness-theory}

We begin by showing that for any fixed non-max range value there exists an instantiation of the agents' movements for which no recharging will take place. 

\begin{prop}\label{prop:bad-fixed-scenario}
For any range value $R < R_{\max}$, there exists a scenario for which fixing the range equal to $R$ is equivalent to not using a charger at all.
\end{prop}

\begin{proof}
Consider the mobility scenario according to which no agent ever passes through the circle $\calC_R$. Then, if the range is set to $R$ for the whole period of time, no agent will ever get recharged. 
\end{proof}

A scenario similar to the one in the proof of Proposition~\ref{prop:bad-fixed-scenario} exists even for the maximum possible range $R_{\max}$. However, in such a case there exists {\em no} algorithm that can do any better. Hence, we need to make the critical assumption that all agents will pass through the circle $\calC_{R_{\max}}$ at least once. Next, we prove a stronger statement that holds true even when we consider the maximum range value. 

\begin{prop}\label{prop:bad-max-scenario}
There exists a scenario for which setting the charger's range equal to any fixed value $R$ is not optimal.
\end{prop}

\begin{proof}
Consider the scenario according to which the agents get in range only when their energy levels are below a threshold. This scenario captures cases where the agents correspond to humans using smart devices; they recharge their devices only when they need to. 
Assume that the agents have the following energy consumption characteristics. There are $n-1$ agents with small energy consumption and a {\em single} greedy agent that consumes all of its available energy, at every round. 

If the charger's range is fixed to any $R$ during the whole time horizon, this single greedy agent can choose its in-range position so that it gets its battery fully recharged. As a result, the charger's energy can be quickly drained out (if the initial energy is small enough), before the other agents have a chance to get recharged. In contrast, consider the algorithm that adapts to the behavior of this greedy agent and, in each round, sets the range such that this agent gets a minimum amount of energy. For example, it can set the range equal to the distance between the agent and the charger so that, according to equation (\ref{eq:received-energy}), it gives to the agent only a small amount of energy every time. This way, the charger conserves energy for the rest of the agents and the network's lifetime can be expanded.  
\end{proof}

\subsection{Experimental justification}\label{sec:need-for-adaptiveness-experiments}
To experimentally demonstrate the phenomenons that were observed in the previous subsection, we will now compare two fixed value algorithms and an adaptive one using a simulation setup with $n=100$ agents that move around in a $25 \times 25$ network area $\calA$ following a particular mobility model over a time horizon of $T=2500$ rounds.

Let $v_{\max}=3$ be the maximum speed that any agent may have. At the beginning of each round $t$, every agent $i$ has a {\em speed mode} $\mu_i(t) \in [3]$, which is redefined with probability $1/4$ (otherwise, it remains the same). The speed mode indicates whether the velocity of the agent takes random values in the interval $I_1 = [0,\frac{1}{4}v_{\max}]$, $I_2 = \left( \frac{1}{4}v_{\max},\frac{1}{2}v_{\max} \right]$, or $I_3 = \left(\frac{1}{2}v_{\max},v_{\max} \right]$, and aims to model three kinds of movement: slow (like walking), medium (like running), and fast (like travelling in a vehicle). Each agent $i$ performs a {\em random walk} as follows. At round $t$, it starts from position $p_i(t) \in \calA$, and randomly chooses a new direction $\theta_i(t) \in [0,2\pi)$ and a new velocity $v_i(t) \in I_{\mu_i(t)}$. The direction $\theta_i(t)$ together with $p_i(t)$, define a line along which the agent travels with the chosen velocity $v_i(t)$ until it reaches its final position at the end of the round, which is the position $p_i(t+1) \in \calA$ at the beginning of the next round. In particular, $p_i(t+1)$ has coordinates 
$$x_i(t+1) = x_i(t) + v_i(t) \cdot \tau \cdot \cos{\theta_i(t)}$$ 
and
$$y_i(t+1) = y_i(t) + v_i(t) \cdot \tau \cdot \sin{\theta_i(t)}.$$
If these equations do not define a point in $\calA$, then the movement is redefined accordingly. Starting from $t=1$ and the initial deployment of the agents in $\calA$, the above process is repeated for all rounds $t \in [T]$.
\footnote{Notice that the mobility model we consider here is similar to the random way-point model, but we also allow for special restrictions in the movements of the agents that give birth to many interesting and extreme scenarios like the ones identified in Section~\ref{sec:need-for-adaptiveness-theory}; this way we can also partially avoid some harmful properties of the random way-point model (for instance, convergence of the agents towards the center of the network area, where the charger is positioned~\citep{YLN03}). Let us remark here that we do utilize these worst-case scenarios in our experimental evaluation in Section~\ref{sec:algorithms}, where we consider probability distributions over both general and special mobility scenarios to test our algorithms in highly heterogeneous settings.}

The charger is positioned at the center of $\calA$, has initial energy $C=10^5$, and its range can take values in the interval $[1,5]$. All agents have battery $B=1000$ and are randomly partitioned into 4 groups, namely, $(S_1, S_2, S_3, S_4)$ of expected sizes $\left(\frac{n}{2}, \frac{n}{4}, \frac{n}{8},\frac{n}{8} \right)$. Then, agent $i$ consumes energy following a poisson distribution with randomly chosen expected value $\gamma_i$ such that 
$\gamma_i \in [0,10 \cdot 2^{j-1}]$ if $i \in S_j$.
We remark that the expected values are chosen non-uniformly from the corresponding intervals so that there is heterogeneous energy consumption among the agents.

The first fixed value algorithm sets the range equal to $\frac{1}{2}(R_{\min} + R_{\max}) = 3$ during the whole period of time, while the second one sets the range equal to $R_{\max}=5$; to simplify our discussion, in the following we will refer to these as the $3$- and $5$-algorithm, respectively. The adaptive algorithm is simple and oblivious to the characteristics of the agents: at the beginning of each round, it equiprobably sets the range equal to $1$ or $5$; even thought this adaptive algorithm is not really sophisticated, its randomized nature allows it to escape from problematic mobility scenarios.

Furthermore, we also compare these algorithms to the optimal one when the charger is given {\em infinite} energy. This optimal algorithm of course sets the range equal to the maximum possible during any round. Its performance serves as an upper bound that is {\em unreachable} by any algorithm when the charger has finite energy.

We present results for two different setups corresponding to two different mobility scenarios. In the first one, all agents randomly move around the whole network area. In the second one, no agent is allowed to pass through the circle $\calC_3$. The first scenario aims to capture random movements, while the second one follows Proposition~\ref{prop:bad-fixed-scenario} and serves as an extreme case for small range values. Recall that we would like our algorithms to perform efficiently in both scenarios, as the agents' characteristics are generally unknown and become partially available in an online manner. 

\begin{figure*}[p]
\centering
\begin{subfigure}{0.46\textwidth}
   \includegraphics[width=0.9\columnwidth]{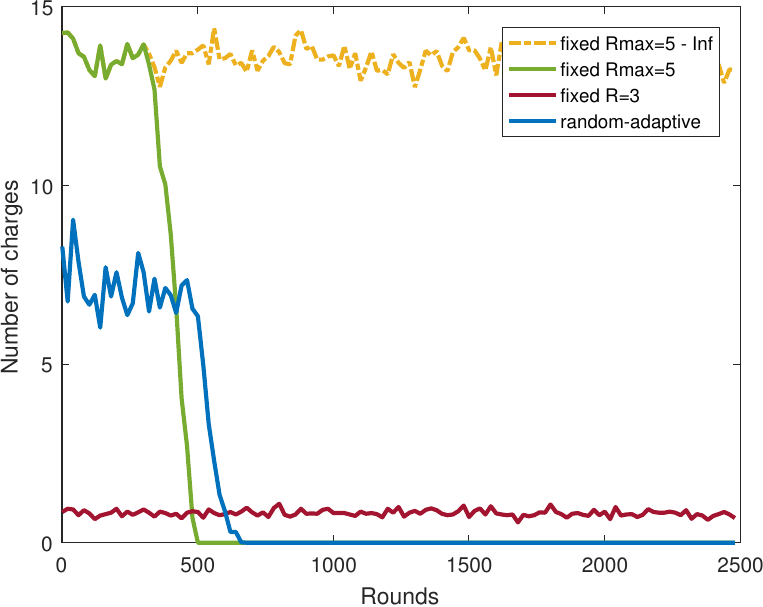}
   \caption{}
   \label{fig:adaptive-vs-fixed-a}
\end{subfigure}
\begin{subfigure}{0.46\textwidth}
   \includegraphics[width=0.9\columnwidth]{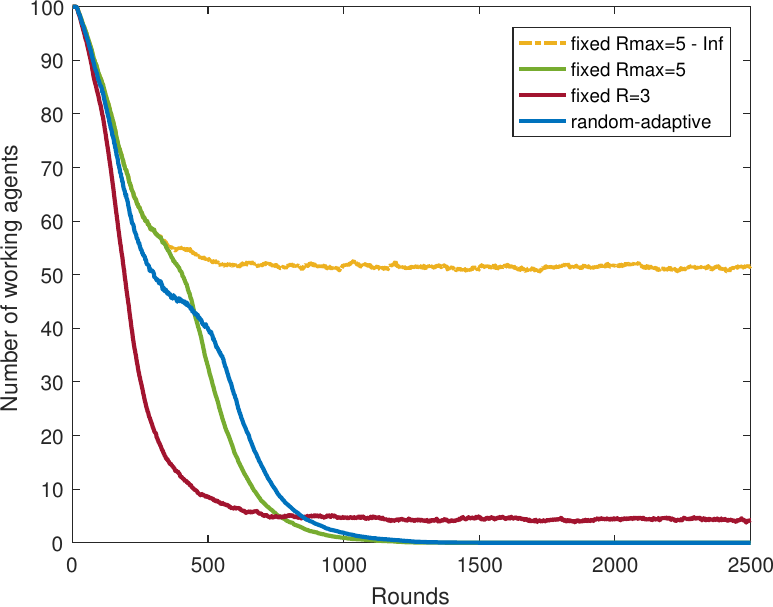}
   \caption{}
   \label{fig:adaptive-vs-fixed-b}
\end{subfigure}

\begin{subfigure}{0.46\textwidth}
   \includegraphics[width=0.9\columnwidth]{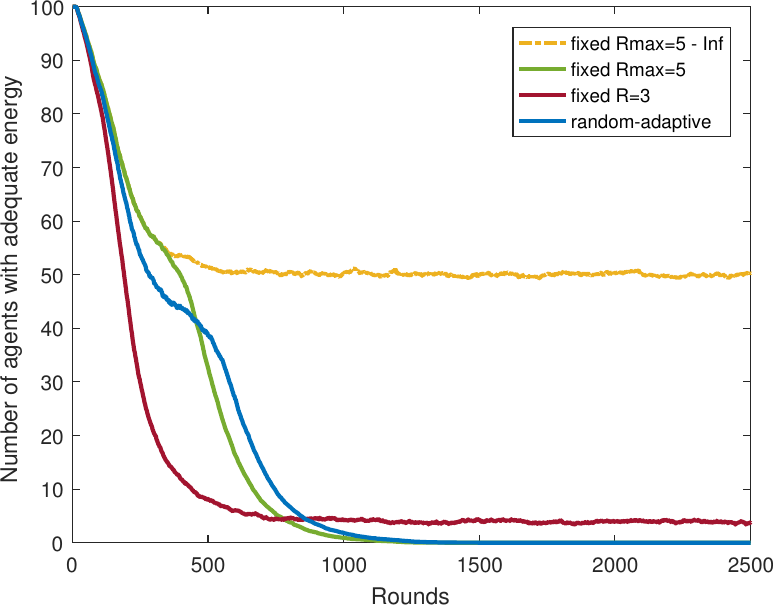}
   \caption{}
   \label{fig:adaptive-vs-fixed-c}
\end{subfigure}
\begin{subfigure}{0.46\textwidth}
   \includegraphics[width=0.9\columnwidth]{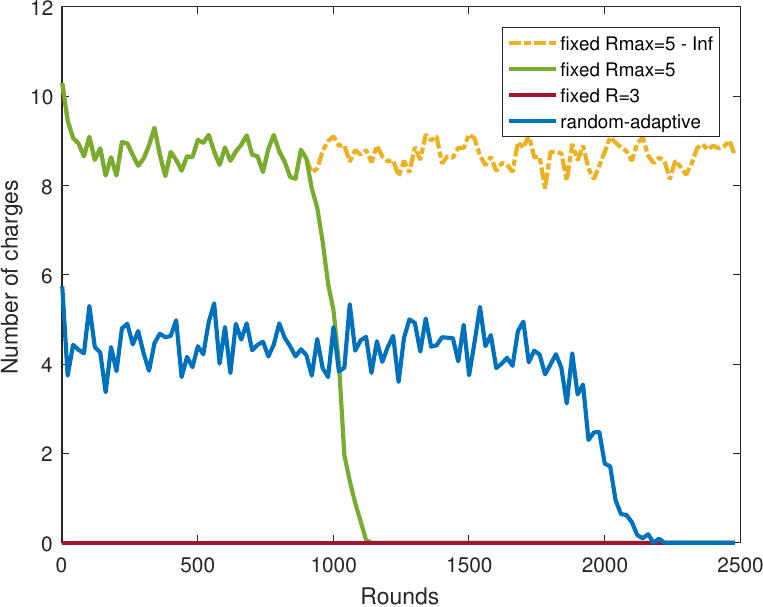}
   \caption{}
   \label{fig:adaptive-vs-fixed-d}
\end{subfigure}

\begin{subfigure}{0.46\textwidth}
   \includegraphics[width=0.9\columnwidth]{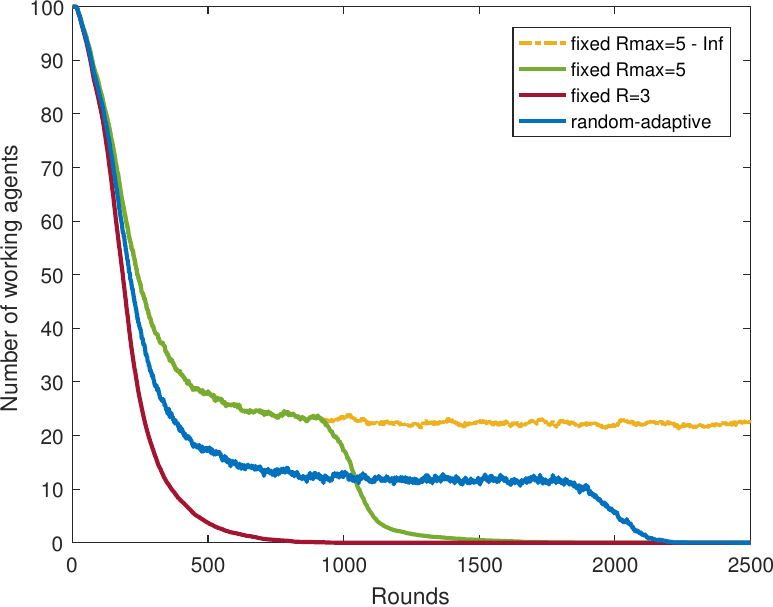}
   \caption{}
   \label{fig:adaptive-vs-fixed-e}
\end{subfigure}
\begin{subfigure}{0.46\textwidth}
   \includegraphics[width=0.9\columnwidth]{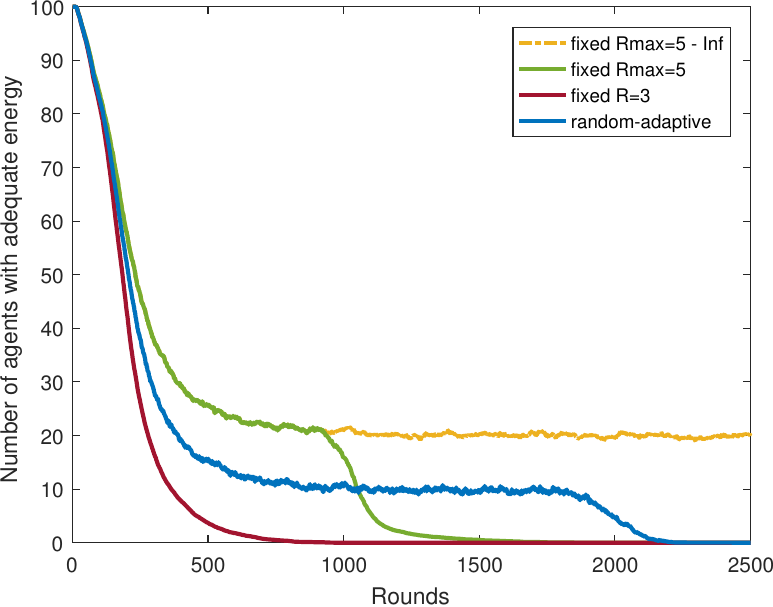}
   \caption{}
   \label{fig:adaptive-vs-fixed-f}
\end{subfigure}
\caption{Comparison between two fixed value algorithms and the randomized adaptive algorithm that chooses between the minimum and the maximum range equiprobably at each round. Figures (a)--(c) correspond to the scenario where all agents randomly move around the network area, while Figures (d)--(f) correspond to the scenario where the agents are not allowed to enter $\calC_3$. Figures (a) and (d) depict the number of charges over time. Figures (b) and (e) depict the number of working agents over time. Finally, Figures (c) and (f) depict the number of agents that have adequate energy to fully complete a communication task during each round. For statistical smoothness, the simulation has been repeated for $100$ times and the depicted lines correspond to average performance.}
\label{fig:adaptive-vs-fixed}
\end{figure*}

Figure~\ref{fig:adaptive-vs-fixed} depicts the performance of the algorithms with respect to three different objectives: 
\begin{itemize}
\item the number of charges (in Figures~\ref{fig:adaptive-vs-fixed-a} and~\ref{fig:adaptive-vs-fixed-d}),
\item the number of {\em working} agents that either have energy at the beginning of a round or get recharged during it (in Figures~\ref{fig:adaptive-vs-fixed-b} and~\ref{fig:adaptive-vs-fixed-e}), and
\item the number of agents with {\em adequate} energy to complete their communication tasks during a round (in Figures~\ref{fig:adaptive-vs-fixed-c} and~\ref{fig:adaptive-vs-fixed-f}).
\end{itemize}
The third objective (number of agents with adequate energy) is stronger than the second one (number of working agents), and the fact that the corresponding figures are very similar indicates that the quality of the recharges is sufficient.

As expected, in both simulations, the $5$-algorithm recharges more agents during the early rounds, essentially simulating the infinite-energy optimal algorithm. However, since the charger's energy is finite, it is drained out quickly. On the other hand, the $3$-algorithm consistently recharges less agents but over a longer period of time in the first simulation, while it performs poorly and is equivalent to not having a charger (zero charges) in the second simulation. The adaptive algorithm performs sufficiently in the first simulation where it strikes a balance between the two fixed value algorithms, while it outperforms both of them in terms of keeping the network active for longer time in the second simulation. Notice the difference between the $3$-algorithm and the adaptive one, even though the expected range of the latter is exactly equal to $3$. 
  
\section{Optimization problems}\label{sec:hardness}
In this section, we define two simplified offline optimization problems and prove their computational intractability. These two problems are closely related to the online one, which we defined in the previous sections, and each of them focuses on optimizing a particular objective goal. Specifically, the first objective is to maximize the number of charges that the charger performs during a given time horizon, which naturally quantifies the working utilization of the charger. The second objective is to maximize the number of rounds during which the network is active; this is of course desirable since the whole point of carefully managing the charging process is in order to be able to extend the lifetime of the network. The hardness of these two optimization problems is only indicative of the hardness of the actual online multi-objective problem. Since the mobility and energy consumption characteristics of the agents are revealed in an online manner, we are missing the appropriate information to optimally compute the most efficient sequence of charging ranges. The hardness of the two problems indicates that even if we did have this information, there will always exist instances that cannot be quickly solved.

As input, we are given all information about the behavior of the agents during a time horizon $T$. The charger has initial energy $C$ and its range can be chosen from a set of $k$ distinct values $\{R_1, ..., R_k\}$ such that $0 \leq R_1 < ... < R_k$. All non-fully charged agents that are in the specified charging range receive appropriate energy according to the adopted charging model. For any $t \in [T]$, the objective of MNC (standing for {\sc Maximize Number of Charges}) is to set the range $R(t)$ of the charger in order to maximize the total number of recharged agents until the charger is left out of energy. The objective of MNL (standing for {\sc Maximize Network Lifetime}) is to set the range $R(t)$ in order to maximize the total rounds during which there exists at least one agent with non-zero (strictly positive) energy. 

\begin{thm}\label{thm:MNC-hardness}
MNC is NP-hard, even for two range values.
\end{thm}

\begin{proof}
We use a reduction from the {\sc Knapsack Problem} (KP, for short) which is known to be NP-hard~\citep{GareyJohnson}. Its formal description is as follows.

\begin{quote}
KP: Consider a collection of $q$ items $a_1, ..., a_q$ such that item $a_i$ has value $v(a_i) \in \R_{\geq 0}$ and weight $w(a_i) \in \R_{\geq 0}$. We are given a knapsack of capacity $W \in \R_{\geq 0}$, and the goal is to select a set of items of total weight at most $W \in \R_{\geq 0}$ in order to maximize the total value of these items.
\end{quote}

Given an instance of KP we will design an instance of MNC. First, without loss of generality, we assume that the values of the items as well as the weight $W$ of the knapsack in the instance of KP are rescaled so that they are integer numbers (for example they are all multiplied by some large number). Second, there are no items with zero value (as such items can be discarded) and no items with zero weight (as such items are for free). 

Now, our MNC instance is as follows:
\begin{itemize}
\item There are $n = \max_t v(a_t)$ agents with battery $B = \max_t \frac{w(a_t)}{v(a_t)}$. 
\item The initial energy of the charger is $C = W$ (the knapsack corresponds to the charger).
\item There are $T = q$ rounds (every item corresponds to a round) and each of them lasts for a unit of time.
\item The range of the charger can either be set to $0$ or $R = \max_t \sqrt{ \frac{w(a_t)}{v(a_t)} }$; essentially, the charger is either {\em inactive} or {\em active} (and its range is $R$). 
\item For each round $t$, the movement and energy consumption characteristics of the agents are as follows. At the beginning of the round, all agents are fully charged. There is a set $A_t$ of exactly $v(a_t)$ agents at distance $d_t = R \sqrt{ \frac{v(a_t)}{w(a_t)} }$ each of whom travels along the circle $\calC_{d_t}$, and consumes energy equal to $\frac{w(a_t)}{v(a_t)} \leq B$ in case the charger is active, and $0$ otherwise; such an energy consumption may be due to the communication of the agents with the charger itself. All other agents (if there are any) do not have any energy consumption during round $t$ and move arbitrarily (but consistently to future positioning requirements).
\end{itemize}

Now, let us focus on an arbitrary round $t \in [q]$. If the charger is inactive during this round, then of course no agent gets recharged. However, according to the above specified energy consumption characteristics, all agents remain fully charged in such a case. On the other hand, if the charger is active during round $t$, then according to equation (\ref{eq:received-energy}) with $\alpha=1$ and $\beta=0$, every agent in $A_t$ receives energy equal to 
$$\frac{R^2}{d_t^2} = \frac{R^2}{R^2 \frac{v(a_t)}{w(a_t)}} = \frac{w(a_t)}{v(a_t)},$$
which is exactly its energy consumption during this round. Therefore, the charger needs to spend $w(a_t)$ units of energy in total in order to fully recharge these $v(a_t)$ agents during round $t$. In other words, the number of charges corresponds to the total value of the selected items and the total needed energy corresponds to the total weight of these items. Consequently, any set of items with maximum total value satisfying the knapsack capacity corresponds to a set of rounds during which the charger is active with maximum number of charges satisfying the initial energy of the charger, and vice versa. The proof is complete.  
\end{proof}

\begin{thm}\label{thm:MNL-hardness}
MNL is NP-hard, even for one agent and two range values.
\end{thm}

\begin{proof}
We again use a reduction from KP (see the proof of Theorem~\ref{thm:MNC-hardness} for its formal definition). Given an instance of KP, we define an instance of MNL:
\begin{itemize}
\item There is a single agent with battery $B = \max_i w(a_i)$.
\item The initial energy of the charger is $C = W$ (the knapsack corresponds to the charger).
\item Every round lasts for a unit of time.
\item The charger can either be inactive with zero range or active with range $R = \max_i \sqrt{ \frac{w(a_i)}{v(a_i)} }$.
\item During the first round $t_1$, the agent is out of the range of the charger and consumes all of its battery. For each item $i \in [q]$, there is time horizon $T_i$ consisting of $v(a_i)$ rounds. During the first of these rounds the agent is in range at fixed distance $d_i = R\sqrt{\frac{1}{w(a_i)}}$ (for example it travels along the circle $\calC_{d_i}$ or is static), while for the remaining $v(a_i)-1$ rounds, the agent moves out of range and has a total energy consumption of $w(a_i) \leq B$ so that during all these rounds it has non-zero energy. These $q$ time horizons are continuous, given a permutation of the items: $T = t_1 \cup_i T_i$.
\end{itemize}

If the charger is inactive during the first round of any time horizon $T_i$, then the agent does not get and does not have any energy during $T_i$ (a total of $v(a_i)$ rounds). On the other hand, if the charger is active during the first round of $T_i$, since the agent is at distance $d_i$ from the charger, and using equation (\ref{eq:received-energy}) with $\alpha = 1$ and $\beta=0$, the energy that the agent receives by the charger is equal to 
\begin{align*}
\frac{R^2}{d_i^2} = \frac{R^2}{R^2 \frac{1}{w(a_i)}} = w(a_i),
\end{align*}
which is exactly the energy that it consumes during $T_i$. Therefore, if the charger is active during the first round of $T_i$, the agent is active for $v(a_i)$ rounds and the charger spends $w(a_i)$ units of energy. As a result, the number of rounds that the agent is active is equal to the total value of the selected items. Hence, any set of items with maximum total value satisfying the knapsack capacity corresponds to a set of time horizons with maximum number of rounds (during which the agent is active) satisfying the energy capacity of the charger, and vice versa. The proof is complete.  
\end{proof}

\section{Maximizing the network lifetime: the case of one agent and on-off charger}
Here, we consider the very restricted version of the optimization MNL problem with a single agent, while the charger can be either active (with some charging range) or inactive (with zero range); we refer to this restriction as 1-MNLb. We next show that 1-MNLb can be reduced to the Knapsack problem (KP).
 
\begin{thm}\label{thm:1-MNLb-reduction}
The 1-MNLb problem reduces to KP.
\end{thm}

\begin{proof}
Consider an instance of 1-MNLb with time horizon $T$ and active range equal to $R>0$. We construct the following instance of KP:
\begin{itemize}
\item The capacity of the knapsack is equal to the initial energy of the charger: $W = C$.
\item For every round $t \in [|T|]$, there is an item $\alpha_t$ with value $v(\alpha_t)=1$ and weight $w(\alpha_t)$ that depends on the position and energy of the single agent $x$:
\begin{align*}
w(\alpha_t) =
\begin{cases}
\frac{\alpha \cdot R^2 \cdot T_x^\inside(t)}{(||p_{\charger}-f_x(t)|| + \beta)^2}, & \text{ if } f_x(t) \text{ is defined } \\
W+\epsilon, & \text{ if } f_x(t) \text{ not defined and } E_x(t) = 0 \\
0, & \text{ if } f_x(t) \text{ not defined and } E_x(t) > 0
\end{cases}
\end{align*}
\end{itemize}
Observe that if the agent does not enter the circle $\calC_R$ or the charger has range $R(t)=0$ during round $t$, then $f_x(t)$ is not defined.
\begin{itemize}
\item If the agent has zero energy, then the item $\alpha_t$ has weight that exceeds the capacity of the knapsack and, therefore, it cannot be selected in any solution of the KP instance.
\item If the agent has strictly positive energy, the weight of item $\alpha_t$ is zero and it can always be selected in the KP instance since it is for free. 
\end{itemize}
In the case where the agent enters the circle $\calC_R$ and the charging range is $R(t)=R$, then $f_x(t)$ is defined and the weight of the corresponding item $\alpha_t$ is exactly equal to the energy that the agent will receive by the charger. Therefore, the decision of setting $R(t)=R$ at some round $t$ during which the agent enters the circle $\calC_R$ is equivalent to the decision of selecting the corresponding item $\alpha_t$ to the knapsack instance. We can now conclude that any charging sequence that maximizes the number of rounds during which the agent has strictly positive energy defines a feasible selection of items (corresponding to rounds during which the agent has positive energy -- either due to residual energy or due to recharging) of maximum value for the KP instance, and vice versa.
\end{proof}

Due to Theorems \ref{thm:MNL-hardness} and \ref{thm:1-MNLb-reduction}, and the fact that KP admits a fully polynomial time approximation scheme~\citep{V01}, we obtain the following statement.

\begin{cor}
There exists an FPTAS for  the 1-MNLb problem.
\end{cor} 

Of course, the existence of such a strong approximation algorithm for the 1-MNLb problem is of theoretical interest only. Besides its specialization on the case of one agent, it requires full knowledge of the characteristics of the agent throughout the time horizon, which is unrealistic.

\section{Adaptive algorithms and experimental evaluation}\label{sec:algorithms}
We propose three adaptive (heuristic) algorithms and experimentally compare them to each other. The algorithms are presented in an increasing order in terms of the knowledge they require in order to decide the charging range during any round $t$. The first algorithm uses information about the position $p_i(t)$ of every agent $i$ for whom it is $p_i(t) \in \calC_{R_{\max}}$. The other two algorithms require information about the positions $p_i(t)$ and $p_i(t+1)$ as well as the energy level $E_i(t)$ of every agent $i$ in $\calA$. Moreover, the third algorithm needs additional information about the energy consumption of the agents.\footnote{Here, we implicitly assume that the communication range of the charger is equal to the maximum possible charging range. Therefore, all agents that are within the maximum range at the beginning of a round can simultaneously send all necessary information (position, direction, speed, current energy level, and energy that will be consumed during the round) to the charger. Of course, such a communication process can be performed efficiently only for small-scale networks, where the agents are inherently quite close to the charger. See the discussion in Section~\ref{sec:conclusion} for generalizations of our model that could be considered for networks of larger scale.} As one can see by their definitions below, the algorithms also differ substantially in their computational complexity as well. 

\paragraph{Least Distant Agent or Maximum Range (LdMax)}
The LdMax algorithm uses a parameter $q \in [0,1]$ and works as follows. At the beginning of each round $t$,  it sets 
\begin{align*}
R(t) := 
\begin{cases}
\max\{R_{\min},{\min}_{i: p_i(t) \in \calC_{R_{\max}}} ||p_\charger-p_i(t)|| \}, & \text{with probability } q \\
R_{\max}, & \text{otherwise}.
\end{cases}
\end{align*}
This is a generalization of the randomized algorithm that we considered in Section~\ref{sec:need-for-adaptiveness} which sets the range equiprobably to $R_{\min}$ or $R_{\max}$. The difference here is that there is a probability of setting the range equal to the distance between the charger and its closest agent (if this is a valid range value) in order to capture worst-case scenarios where there are no agents close to the charger.  

\paragraph{Maintain Working Agents (MWA)}
The MWA algorithm uses a parameter $\mu \in [n]$ and, during each round $t$, sets the range $R(t)$ in an attempt to guarantee that there are at least $\mu$ working agents in the network (i.e. agents that either have positive energy at the beginning of the round or get recharged during it).
To find the appropriate range $R(t)$ it works as follows. First, it counts the number $k_1(t)$ of  agents that are in $\calC_{R_{\max}}$ and have positive energy at the beginning of the round. If $k_1(t) \geq \mu$, then it sets $R(t):=R_{\min}$ since the requirement is already satisfied. Otherwise, it counts the number $k_2(t)$ of agents that have zero energy at the beginning of the round and $p_i(t) \in \calC_{R_{\max}}$ or $p_i(t+1) \in \calC_{R_{\max}}$. If $k_1(t) + k_2(t) < \mu$, then it sets $R(t):=R_{\max}$ since the requirement cannot be satisfied. Otherwise, it searches for the smallest $R^*$ such that the circle $\calC_{R^*}$ covers at least $\mu-k_1(t)$ agents, and sets $R(t) := R^*$.

\paragraph{Maximize Charges over Energy Ratio (MCER)}
Let $\calR$ be a set of discrete range values in $[R_{\min},R_{\max}]$. Let $\nu_j(t)$ be the number of agents that get recharged when the range is equal to $R_j \in \calR$ during round $t$, and let $\varepsilon_j(t)$ be the total given energy in this case. The MCER algorithm uses a parameter $\lambda \geq 1$ and sets 
$$R(t) := \underset{R_j \in \calR}{\arg\max}\frac{\nu_j(t)^\lambda}{\varepsilon_j(t)}.$$
This algorithm is inspired by a simple greedy $2$-approximation algorithm for the Knapsack problem and attempts to strike a balance between the number of charges and the energy that it has to give in order to perform these charges. However, observe that it needs to perform many heavy computations as, in order to choose the best range, it has to simulate the whole recharging process multiple times. 

\subsection{Simulation Setup}
We now experimentally compare these adaptive algorithms. We consider a simulation setup similar to the one we considered in Section~\ref{sec:need-for-adaptiveness-experiments}.\footnote{We remark that the setup that we present here is only indicative. Actually, we have experimented with many different setups that differ on the number of agents and their battery capacity, the network size, and the initial energy of the charger. For all such setups, the relative performance of our algorithms is similar.} There are $n=100$ agents that move around in a $25 \times 25$ network area $\calA$ following the mobility model described in Section~\ref{sec:need-for-adaptiveness-experiments} such that each agent has $v_{\max} = 3$ and speed mode that is redefined with probability $1/4$ in each round. The charger is positioned at the center of $\calA$, has initial energy $C=10^5$, and its range can take values in $[1,5]$. All agents have battery $B=1000$ and are randomly partitioned into 4 groups $(S_1, S_2, S_3, S_4)$ of expected sizes $\left(\frac{n}{2}, \frac{n}{4}, \frac{n}{8},\frac{n}{8} \right)$. Then, agent $i$ consumes energy following a poisson distribution with randomly chosen expected value $\gamma_i$ such that $\gamma_i \in [0,10 \cdot 2^{j-1}]$ if $i \in S_j$. 

For the agent mobility behavior we consider three randomized scenarios:
\begin{itemize}
\item (S1) All agents randomly move around in $\calA$;
\item (S2) Choose $R \in [R_{\min}, \frac{1}{2}R_{\max}]$ uniformly at random so that no agent is allowed to enter circle $\calC_R$;
\item  (S3) Choose $\delta \in \left[ \lfloor\frac{n}{10}\rfloor \right]$, $R_\ell \in \left[ R_{\min}, \frac{1}{4}(R_{\min}+R_{\max}) \right)$ and $R_h \in \left[ \frac{1}{4}(R_{\min}+R_{\max}), R_{\max} \right]$ uniformly at random so that $\delta$ agents live in the ring $\calC_{R_h} \setminus \calC_{R_\ell}$, while the remaining $(n-\delta)$ agents randomly move around in $\calA$. 
\end{itemize}
We create a probability distribution over these three mobility scenarios by repeating our simulation for $100$ times so that a different scenario is chosen equiprobably every time. Observe that there are many different random choices to be made and these give birth to many different instantiations. The goal is to test our algorithms under a highly heterogeneous setting.

\begin{figure*}[p]
\centering
\begin{subfigure}{0.46\textwidth}
   \includegraphics[width=0.9\columnwidth]{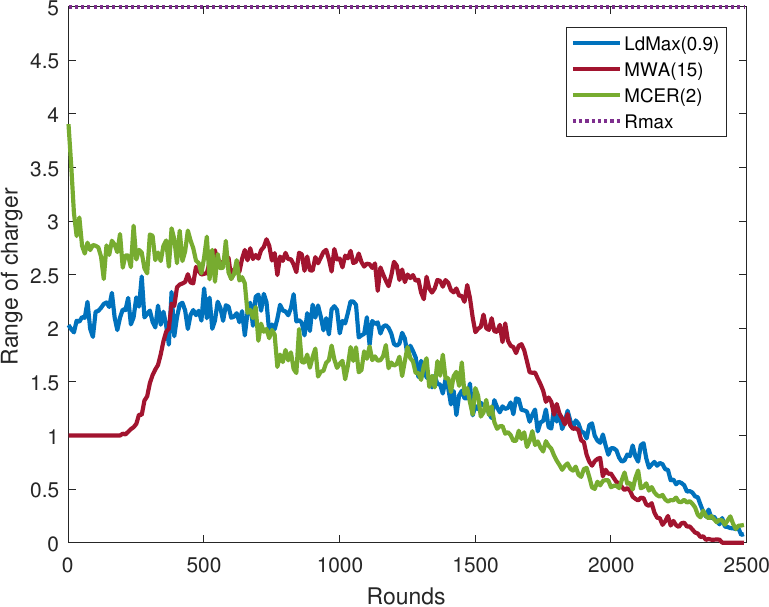}
   \caption{}
   \label{fig:adaptive-range}
\end{subfigure}
\begin{subfigure}{0.46\textwidth}
   \includegraphics[width=0.9\columnwidth]{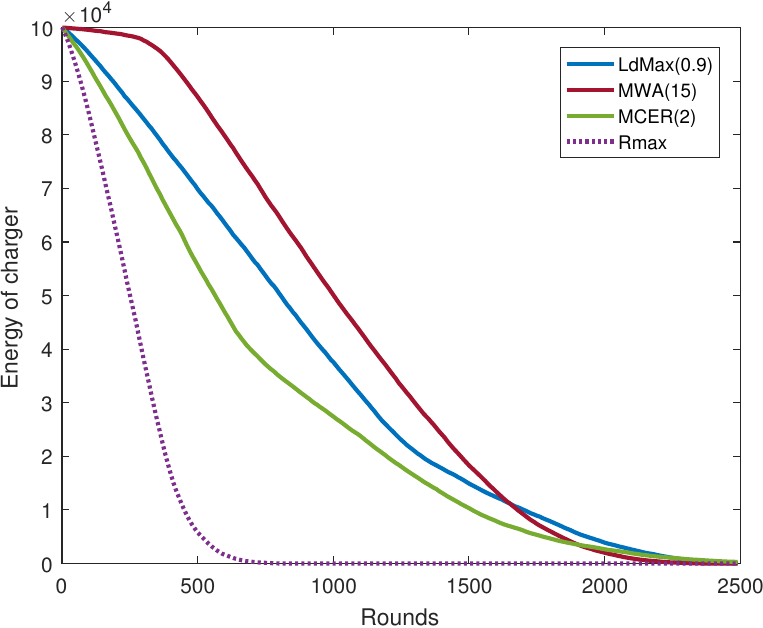}
   \caption{}
   \label{fig:adaptive-ch-energy}
\end{subfigure}

\begin{subfigure}{0.46\textwidth}
   \includegraphics[width=0.9\columnwidth]{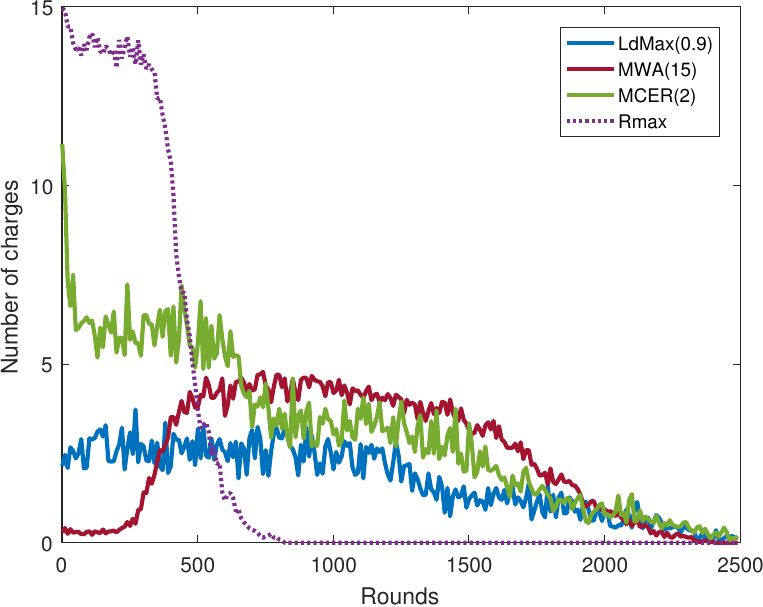}
   \caption{}
    \label{fig:adaptive-charges}
\end{subfigure}
\begin{subfigure}{0.46\textwidth}
   \includegraphics[width=0.9\columnwidth]{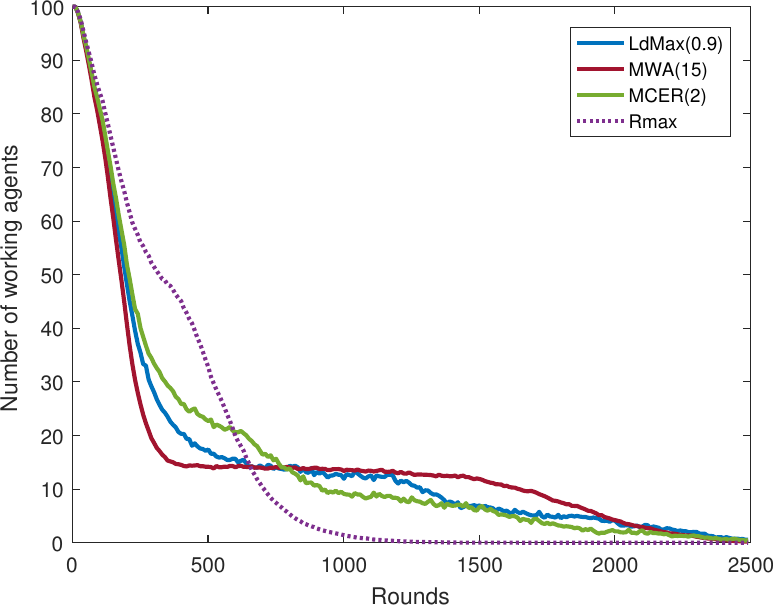}
   \caption{}
   \label{fig:adaptive-working}
\end{subfigure}

\begin{subfigure}{0.46\textwidth}
   \includegraphics[width=0.9\columnwidth]{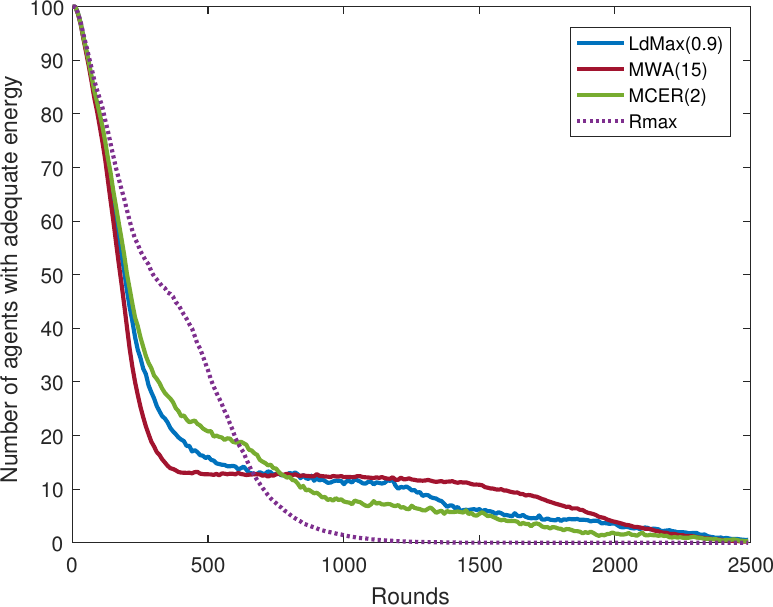}
   \caption{}
    \label{fig:adaptive-adequate}
\end{subfigure}
\begin{subfigure}{0.46\textwidth}
   \includegraphics[width=0.9\columnwidth]{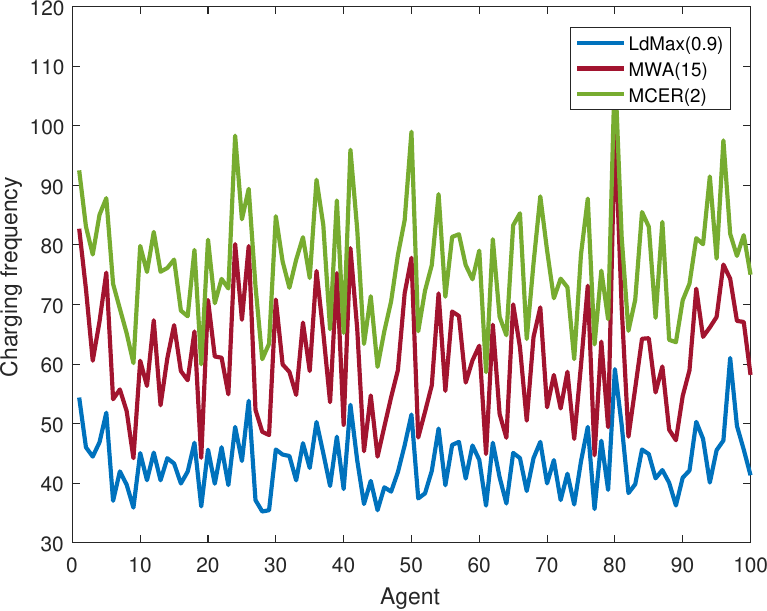}
   \caption{}
   \label{fig:adaptive-frequency}
\end{subfigure}

\caption{Comparison between the three adaptive algorithms LdMax(0.9), MWA(15) and MCER(2) as well as the fixed  $R_{\max}$ value algorithm. Figure (a) depicts the evolution of the charging range over time. Figure (b) depicts the decrease of the charger's energy over time. Figure (c) depicts the number of charges that were performed over time. Figure (d) depicts the number of working agents over time. Figure (e) depicts the number of agents with adequate energy over time. Figure (f) depicts the charging frequency of the agents (the number of times they were recharged). The simulated data presented here are averages over $100$ executions. The performance of each algorithm in the different executions is robust and sharply concentrated around the depicted average value.}
\label{fig:adaptive}
\end{figure*}

\subsection{Results and interpretation}
After extensive fine-tuning of the parameters used by our adaptive algorithms, we have concluded that setting $q=0.9$, $\mu=15$ and $\lambda=2$ are the best values for the particular simulation setup that we consider here. In general, we expect $q$ to depend heavily on the density of the network; it should be smaller for more sparse networks. On the other hand, $\lambda=2$ seems to nicely balance the ratio considered by MCER due to the fact that the given energy is of square order according to equation (\ref{eq:received-energy}). Finally, parameter $\mu$ can be picked by the designer to maintain a sufficient number of agents, depending on the needs of the network, the energy of the charger, etc. 

Figure~\ref{fig:adaptive} depicts the performance of the adaptive algorithms as well as that of the fixed $R_{\max}$ value algorithm over time, with respect to various metrics: 
\begin{itemize}
\item the charger's energy (Figure~\ref{fig:adaptive-ch-energy});
\item the charging range (Figure~\ref{fig:adaptive-range});
\item the number of charges (Figure~\ref{fig:adaptive-charges});
\item the number of working agents (Figure~\ref{fig:adaptive-working});
\item the number of agents with adequate energy (Figure~\ref{fig:adaptive-adequate});
\item the charging frequency of the agents (Figure~\ref{fig:adaptive-frequency}).
\end{itemize}

Due to its definition, MWA guarantees for a long period of time a stable number of working agents (as well as agents with adequate energy). However, MCER seems to outperform the other two algorithms in terms of the total number of charges and the charging frequency of the agents. Essentially, MWA and MCER work in exactly opposite ways, while LdMax lies somewhere in-between of these two, due to its randomized nature. \footnote{We should remark that we  compare our three adaptive algorithms only to each other, as well as to the fixed max range algorithm, and not with the actual optimal solution. Unfortunately, for our simulation setup, where the charge range can take values in the continuous interval $[1,5]$, the space of possible charging range sequences, over which we need to search in order to compute the optimal one, is extremely huge and requires the solution to instances of intractable problems; this would be true even if we considered much simpler setups (like restricting the charging range to take only two possible values).}

To interpret this data, we will briefly analyze how MWA and MCER respond to the behavior of the agents by inspecting Figure~\ref{fig:adaptive-range} which displays the evolution of the charging range over time depending on the algorithm. During the early rounds of the simulation, most of the agents are considered working since they are initially fully charged. Therefore, the requirement of maintaining $15$ working agents is trivially satisfied and MWA starts by having the minimum possible range, so that it stores energy for future use (see Figure~\ref{fig:adaptive-ch-energy}). In contrast, MCER chooses a higher range in order to perform more charges while giving away little energy; since the agents already have energy, they request only a small amount of energy when they get in range, which means that the cost (in energy) per charge is quite small. However, as the time progresses, the energy levels of the agents gradually get lower, there are less working agents, and when an agent gets in range requests for more energy. As a result, MWA is forced to increase the range in order to keep satisfying the requirement of maintaining $15$ working agents, while MCER decreases its range as the cost per charge has increased substantially. 

\subsection{Some scalability issues}
\begin{figure*}[t]
\centering
\begin{subfigure}{0.4\textwidth}
   \includegraphics[width=1\columnwidth]{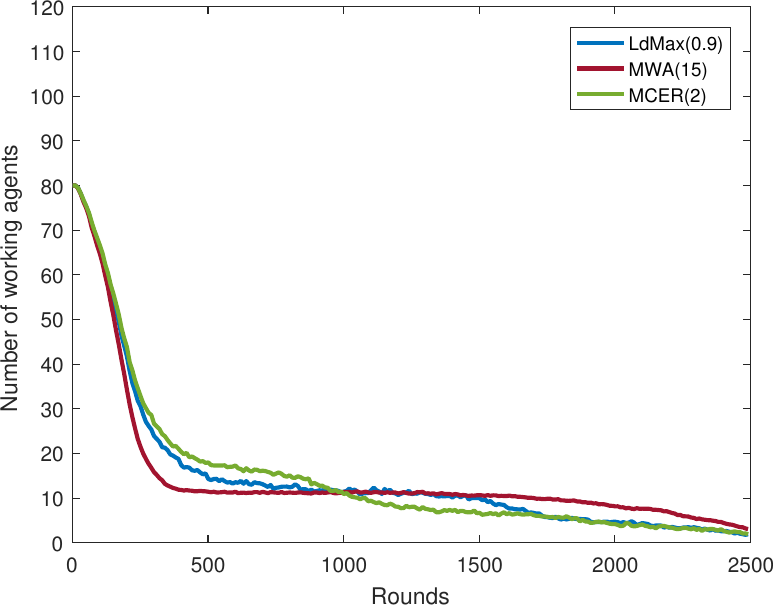}
   \caption{}
\end{subfigure}
\ \ \ \ \ \
\begin{subfigure}{0.4\textwidth}
   \includegraphics[width=1\columnwidth]{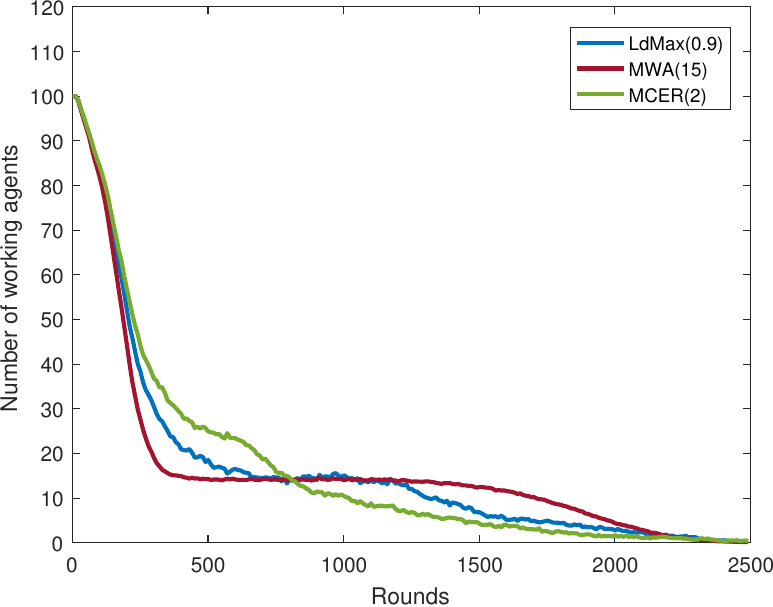}
   \caption{}
\end{subfigure}

\begin{subfigure}{0.4\textwidth}
   \includegraphics[width=1\columnwidth]{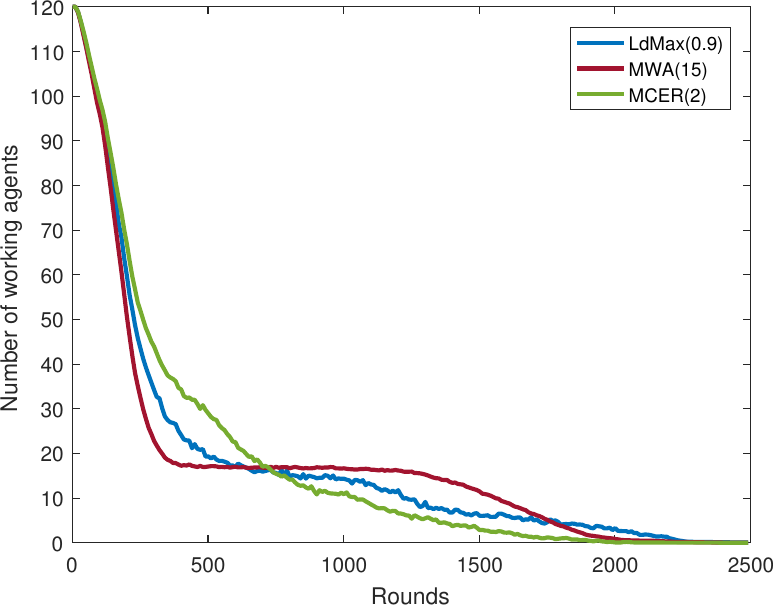}
   \caption{}
\end{subfigure}
\caption{Comparison of the adaptive algorithms LdMax(0.9), MWA(15) and MCER(2) in terms of the number of working agents when there are (a) $80$, (b) $100$, and (c) $120$ agents. Observe that in case (a) the network is more sparse than in case (b), while in case (c) it is even more dense. As a result, there are less and more agents passing through the range of the charger, respectively. This has an analogous impact on the number of working agents and the lifetime of the network as depicted in the three figures. }
\label{fig:different-number-of-agents}
\end{figure*}

We have also experimented with many different values for the number of agents, their battery, as well as the initial energy of the charger. Our results are scalable in the sense that these parameters seem to affect only the network lifetime (it is either increased or decreased) and not the relative performance of the algorithms. Indicatively, Figure~\ref{fig:different-number-of-agents} showcases the performance of our adaptive algorithms, in terms of the number of working agents, when there are $80$, $100$ and $120$ agents, respectively. We remark that, by keeping the network area size fixed and changing the number of agents, we essentially create networks of different densities. However, for networks that operate in larger areas, our setting (using a single static adaptive charger) clearly does not scale well and may result in poor charging performance; see the related discussion in the next section.

\section{Conclusion and possible extensions}\label{sec:conclusion}
In this paper, we studied the problem of dynamically selecting the appropriate charging range of a single static charger in order to prolong the lifetime of a network consisting of mobile agents. We showcased the hardness of the problem by proving the intractability of the two related offline optimization problems MNC and MNL, and presented three interesting heuristics that perform fairly well in the simulation setups that we considered. Of course, there are multiple interesting future directions.

Can we design improved adaptive algorithms with provable efficiency guarantees that perform sufficiently well under any possible scenario, regardless of the mobility and energy consumption characteristics of the agents? A first possible way towards this goal could be to focus on simple fixed range algorithms and algorithms that periodically alternate between different fixed range values (either obliviously or by taking into account some of the agent characteristics). Even though we can easily construct particular worst-case scenarios for which such algorithms perform poorly (like the ones considered in Sections~\ref{sec:need-for-adaptiveness} and \ref{sec:hardness}), there might exist families of interesting special cases where the performance of fixed range algorithms is approximately optimal.

Another way to try to tackle the above question, would be to consider a machine learning like approach. In particular, given statistical information (a prior probability distribution) about the behavior of the agents, is it possible to learn the ``correct'' sequence of values for the charging range in order to prolong the network lifetime as much as possible, while maintaining a fair amount of working agents? We remark that our algorithms do not exploit such training information, and function based only on the online behavior of the agents. Therefore, it is natural to expect that there exist heuristics which significantly outperform the ones that we have proposed in this paper; this is especially true given the experimental performance of LdMax, which is an oblivious randomized heuristic.

The setting that we have studied in this paper is simple and fundamental, but also extremely limited, since using a single static charger may result in poor charging performance when the network area is too large. If the distance between the charger and the agents becomes extremely high, then either the charging range has to be high as well or the agents will not get recharged. Therefore, it is important to also study the natural generalization of using multiple adaptive chargers which may be able to move around and scan the network, or even charge each other in a peer-to-peer manner. In a sense, such a setting would couple (in a non-trivial way) as well as generalize our work together with that of \citet{ABER15}, and definitely deserves investigation. It is worth remarking that, in such settings, since the agents may be able to receive energy from multiple chargers simultaneously (in case the ranges of different chargers overlap and the corresponding agent technology allows multiple recharging), besides the mobility and consumption characteristics of the agents, the locations of all chargers can also critically affect the charging range of each charger.

Finally, in this paper we have modeled the amount of energy that the charger can give to an agent using a simplified version of Friis transmission equation, which is one-dimensional (scalar) and implicitly assumes that charging is a binary process: once an agent is out of range, it does not receive any energy. Therefore, it would be extremely interesting to also consider alternative, more realistic, charging models. One such example, which is able to capture superadditive and cancellation effects in the case of multiple chargers, is the vector model that has recently been considered in \citep{KNRR17,NCB15}. 

\bibliographystyle{named}
\bibliography{adaptive_mobile.bib}

\end{document}